# Training Recurrent Neural Networks via Dynamical Trajectory-Based Optimization

Hamid Khodabandehlou, *Student Member, IEEE*, M. Sami Fadali, *Senior Member, IEEE*

*Abstract*—This paper introduces a new method to train recurrent neural networks using dynamical trajectory-based optimization. The optimization method utilizes a projected gradient system (PGS) and a quotient gradient system (QGS) to determine the feasible regions of an optimization problem and search the feasible regions for local minima. By exploring the feasible regions, local minima are identified and the local minimum with the lowest cost is chosen as the global minimum of the optimization problem. Lyapunov theory is used to prove the stability of the local minima and their stability in the presence of measurement errors. Numerical examples show that the new approach provides better results than genetic algorithm and error backpropagation (EBP) trained networks.

## 1. Introduction

Neural networks are non-parametric models for function approximation. To provide appropriate models, neural networks must be trained using a suitable training dataset. Training neural networks is by minimization a cost function defined using the output of the network and measurements from the modeled system (Zhang and Suganthan, 2016). Classical training approaches use derivatives of the cost function to update the weights of the neural network (Gori and Tesi, 1992; Montingy, 2014). Unfortunately, classical approaches do not guarantee finding the global minimum of the optimization problem and are often trapped in a local minimum. Moreover, classical approaches converge slowly and may have poor generalization capability with noisy data (Hornik, 1991).

Researchers have proposed different modifications to the EBP algorithm to improve its performance and reliability (Zhu et. al., 2006). Jaganathan et. al. (Jaganathan and Lewis, 1996) proposed using delta rule weight tuning to train artificial neural networks. They used passivity theory to show that their weight-tuning algorithm yields a passive neural network. Evolutionary algorithms such as genetic algorithm are another approach to train neural networks. They do not rely on calculating derivatives of the cost function and do not get stuck in local minima of the optimization problem (Georgiopoulos, 2011; Blanco et. al., 2001).

Particle swarm optimization (PSO) is another global optimization approach to train artificial neural networks that was shown to have superior performance to classical optimization approaches (Salerno, 1997; Cai et. al., 2010). Evolutionary PSO is a derivative of the PSO algorithm that learns the internal structure of neural network as well as internal weights (Yu et. al., 2007). It has superior performance to the original PSO algorithm. Simulation results show that IPSONet constructs a compact artificial neural network with good generalization capability. Multi-dimensional PSO is another derivative of PSO to obtain the optimal structure of the neural network which has fast convergence rate in lower dimensions (Kiranyaz et. al., 2009). It was shown to perform better than several other general optimization approaches such as basic genetic algorithms and IPSONet (Abbas, 2003). MPANN is a combination of gradient descent and a multi-objective evolutionary algorithm for learning the internal structure and internal weights of artificial neural networks. It has lower training time than gradient based algorithms. PSO-QI combines mechanical quantum concepts with PSO to produce more logical offspring than other evolutionary algorithms and has faster convergence time (Luitel and Venayagamoorthy, 2009).

Other algorithms have been proposed to train recurrent artificial neural networks. Bounding ellipsoid imposes a bound on the neural network output and the real measurements, and trains the network to satisfy the error bound (Yu and Rubio, 2009). The algorithm is a stable alternative to traditional methods (Rubio and Yu, 2007). EBP through time is an improved version of traditional EBP for training recurrent neural networks which is faster than evolutionary algorithms such as genetic algorithm but still suffers from the local optima issue (Werbos, 1998; Stroeve, 1998; Hermans et. al., 2015; Sharma, 2016). Martens and Sutskever (Martens and Sutskever, 2011) used a combination of Hessian free optimization together with a new damping scheme

This paper is based upon work supported by the National Science Foundation [Grant No. IIA-1301726].
H. Khodabandehlou is with the University of Nevada, Reno, NV, 89557 USA (e-mail: hkhodabandehlou@nevada.unr.edu)
M. Sami Fadali is with the University of Nevada, Reno, NV, 89557 USA (e-mail: fadali@unr.edu)

to train recurrent neural networks. Generalized Long-Short Term Memory (Monner and Reggia, 2012) is an adaptation of Long-Short Term Memory (LSTM) that benefits from the efficient local search of LSTM but is applicable to a larger class of recurrent neural networks. Recursive Bayesian Levenberg-Marquardt was used to train the recurrent neural networks for time-series modeling (Miritikani et. al., 2010). The algorithm is numerically stable and has superior performance to traditional methods, such as extended Kalman training and Newton based methods (Miritikani et. al., 2007).

Although other approaches, such as Levenberg-Marquadrt (Fu et. al., 2015), multi agent systems (Ghazikhani et. al., 2011), dark knowledge transfer (Tang et. al. 2016) and quotient gradient system (Khodabandehlou and Fadali, 2017) have been applied to train artificial neural networks, there are other promising optimization approaches in the mathematics literature that can be applied to training neural networks. In this study, we use a dynamical trajectory-based methodology for training the artificial neural networks. The dynamical trajectory-based methodology consist of two systems: a Quotient gradient system (QGS) that finds the components of the feasible region of the optimization problem, and a projected gradient system (PGS) that searches the feasible components for local optimal solutions of the optimization problem. By switching between the PGS and QGS phases, the algorithm is able to find multiple local minima of the optimization problem and finally finds the global optimal or a good suboptimal solution of the optimization problem.

Unlike Newton-based methods, trajectory-based optimization is a global optimization approach and does not suffer the local minimum issue. Furthermore, trajectory-based optimization does not require a huge number of measurements to find the global optimum and does not have user dependent variables. This makes the methodology more desirable than newton based methods whose performance depends on learning rates and initial point, better than approaches such as particle swarm optimization whose performances depends on initial particles and velocities, and better then genetic algorithm whose performance depends on initial population size, mutation and crossover functions.

Although network parameters must remain bounded, training neural networks is usually posed as an unconstrained optimization problem. In this study, we include upper and lower bounds on the neural network parameters with the minimization of the sum of squared errors and pose neural network training as a constrained minimization problem. Then we use trajectory-based methodology to find components of the feasible region and search those components for local minima. The upper and lower bounds can be chosen arbitrarily large to avoid the effect of conservative bounds on the optimal solution.

The remainder of this paper is organized as follows. Section 2 reviews the dynamical trajectory-based methodology used in this paper. Section 3 describes the application of the dynamical trajectory-based methodology to training neural networks. Section 4 establishes the stability of the training methodology, Section 5 presents simulation results and Section 6 is the Conclusion.

Throughout the paper, we use the following notation. Upper case italics denote matrices and vectors are lower case bold italics. The norm $\|.\|$ is the 2-norm for a vector and induced 2-norm for a matrix. The Frobenius norm of a $A$ matrix is denoted by $\|A\|_F$.

## 2. Dynamical trajectory-based methodology

Lee and Chiang (Lee and Chiang, 2004) proposed a dynamical trajectory-based methodology for finding multiple optimal solutions of nonlinear programming problems. This section reviews their work. Consider the following optimization problem

$$\min f(\mathbf{x})$$
$$\text{s.t. } \mathbf{h}(\mathbf{x}) = \mathbf{0} \tag{1}$$

$f(\mathbf{x})$ is assumed in $C^2(R^n, R^m)$ to guarantee the existence of the solution and $\mathbf{h}(\mathbf{x})$ is assumed to be smooth. The more general case includes inequality constraints as

$$\min f(\mathbf{x})$$
$$\text{s.t. } h_i(\mathbf{x}) = 0, \quad i \in I = \{1,2,\ldots,l\} \tag{2}$$
$$g_j(\mathbf{x}) \leq 0, \quad j \in J = \{1,2,\ldots,s\}$$

The inequality constraint can be transformed into equality constraints by introducing positive slack variables such as $g_j(\mathbf{x}) + s_j^2 = 0$, $j \in J$. $\bar{\mathbf{x}}$ is called the Kuhn-Tucker point of (1) with Lagrange-multipliers $\bar{\boldsymbol{\lambda}} = (\lambda_1, \ldots, \lambda_m)$ if it satisfies the Kuhn-Tucker conditions

$$\nabla_x L(\bar{\mathbf{x}}, \bar{\boldsymbol{\lambda}}) = \nabla f(\bar{\mathbf{x}}) + \sum_{i=1}^{m} \bar{\lambda}_i \nabla h_i(\bar{\mathbf{x}}) = 0$$
$$\nabla_\lambda L(\bar{\mathbf{x}}, \bar{\boldsymbol{\lambda}}) = \mathbf{h}(\mathbf{x}) = 0 \tag{3}$$

$L(\mathbf{x}, \boldsymbol{\lambda}) = f(\mathbf{x}) + \sum_{i \in I} \lambda_i h_i(\mathbf{x})$ is the Lagrangian function of the optimization problem. The feasible region is defined as

$$M := \{x \in R^n : \mathbf{h}(x) = 0\} \tag{4}$$

The feasible region can be any closed subset of $R^n$. The feasible region is subject to the following assumptions:

*Assumptions*:
a) (*Regularity*) At each $x \in M$, $\{\nabla h_i(x), i = 1, ..., m\}$ are linearly independent.
b) (*Nondegeneracy*) At each critical point $\bar{x} \in M$, $\ell^T \nabla^2_{xx} L(\bar{x}, \lambda) \ell \neq 0$ for all $\ell \neq 0$ satisfying $\nabla h_i(\bar{x})^T \ell = 0$ for all $i = 1, 2, ..., m$.
c) (*Finiteness and Separating Property*) $f$ has finitely many critical points in $M$ at which it attains different values of $f$.

Assumption 1 is generically true and when $M$ is compact, the optimization problem is structurally stable (Guddat et. al.,1990; Jongen et. al., 1995). Using the regularity condition together with the implicit function theorem, it can be shown that $M$ is a $n - m$ dimensional smooth manifold. $M$ may be disconnected and nonconcave and is given by the union

$$M = \bigcup_i M_i \tag{5}$$

of disjoint connected components $M_i$. The dynamical trajectory-based method has two main components: the projected gradient system (PGS) and the quotient gradient system (QGS). Starting from an arbitrary initial point, the QGS finds a feasible component of the feasible region, then the PGS finds all the local minima in that component. After finding all the local minima in a particular component, we switch to the QGS to escape from that component and reach another component of the feasible region. By repeating this process, the methodology finds multiple local minima of the minimization problem even if they lie in disjoint regions $M_i$.

## 2.1. PGS phase

The PGS is a nonlinear dynamical system whose trajectories can be used to escape from one local optimal solution and move toward another solution in the current component of the feasible region. The projected PGS is defined as

$$\dot{x} = F(x) = -\nabla f_{\text{proj}}(x), \qquad x \in M \tag{6}$$

where $\nabla f_{\text{proj}}(x)$ is the orthogonal projection of $\nabla f(x)$ to the tangent space $T_x M$ of the constraint set $M$ at $x$. When $Dh(x) = \partial h(x)/\partial x$ is nonsingular, then $\nabla f_{\text{proj}}(x) = P_r(x)\nabla f(x)$ where $P_r(x) = \left(I - Dh(x)^T(Dh(x)Dh(x)^T)^{-1}Dh(x)\right) \in R^{n \times n}$ is the positive semidefinite projection matrix for every $x \in M$. Every local optimal solution of (1) is a stable equilibrium point of the PGS and every trajectory of the PGS converges to one of its equilibrium points. Therefore, starting from any initial point in the stability region of its stable equilibrium, we can reach a local optimal solution. Thus, the problem of escaping from one local optimal solution and moving to another one reduces to escaping from the stability region of one stable equilibrium and entering the stability region of another. To do this, PGS starts from a local optimal solution and proceeds in reverse time until it reaches a decomposition point, which is a saddle point. Starting from the decomposition point, the algorithm moves towards another local optimal solution. Consequently, PGS finds the multiple local minima in the connected component of the feasible region. The next step is to move from one connected component of the feasible region to another by invoking the QGS phase.

## 2.2. QGS Phase

To explore all the components of the feasible region, we must be able to reach a connected feasible component then escape from it and reach another one. We use a nonlinear dynamical system whose trajectories can be used to reach and escape from connected components of the feasible region. The nonlinear dynamical system must have stable equilibrium manifolds that are connected feasible components of the feasible region of the optimization problem. The nonlinear system we need is the following quotient gradient system (QGS)

$$\dot{x} = -Dh(x)^T Dh(x) \tag{7}$$

where $Dh(x)$ is the Jacobian of $h$ at $x$. The right hand side of the QGS equation is the gradient of $\|h(x)\|^2/2$. Lee and Chang showed that the feasible components of (1) correspond to stable equilibrium manifolds of the QGS and that every trajectory of the QGS converges to one of its equilibrium manifolds (Lee and Chiang, 2004). Therefore, a connected component of the feasible region is approached by integrating QGS from any point in the stability region of its stable equilibrium manifold.

Similarly to the PGS phase, escaping from a connected feasible component is by integrating the QGS in reverse time until approaching a point in an unstable manifold on the stability boundary of the stable equilibrium manifold. Then we integrate the QGS forward in time from a point close to the unstable equilibrium manifold until we reach another stable equilibrium manifold. By alternating between the PGS and QGS phases, all the local minima of the optimization problem can be found, and the global minimum can be determined.

## 2.3. Finding Decomposition Points

The decomposition point plays a pivotal role in finding local minima because every two adjacent local optimal solutions of (1) are connected through an unstable manifold of a decomposition point. To define and show how to find a decomposition point, we need several definitions.

A trajectory is the solution of the PGS starting from $x \in M$ at $t = 0$ and is denoted by $\phi(., x): R \to M \subset R^n$. $x^* \in M$ is an equilibrium point if $F(x^*) = 0$. The equilibrium point $x^* \in M$ is said to be hyperbolic if the restriction of Jacobian of $F(.)$ at $x^*$ to the tangent space $T_{x^*}M$ has no eigenvalues with zero real part. If the restricted Jacobian of a hyperbolic equilibrium point has exactly $k$ eigenvalues with positive real parts, it is called a type-$k$ equilibrium point. If all the eigenvalues of the restricted Jacobian have negative real parts, the equilibrium point is called stable; otherwise it is unstable. The stable and unstable manifolds of $x^*$ are defined as follows

$$W^s(x^*) = \left\{x \in M: \lim_{t \to \infty} \phi(t, x) = x^*\right\}$$
$$W^u(x^*) = \left\{x \in M: \lim_{t \to -\infty} \phi(t, x) = x^*\right\} \quad (8)$$

The stability region of stable equilibrium $x_s$ is defined as

$$A(x_s) := \{x \in M: \lim_{t \to \infty} \phi(t, x) = x_s\} \quad (9)$$

The quasistability region of stable equilibrium $x_s$ is defined as $A_p(x_s) = \text{int}(\overline{A(x_s)})$. $A(x_s)$ and $A_p(x_s)$ are open, connected and invariant sets relative to the manifold $M$.

A type-one equilibrium point $x_d$ on the quasistability boundary $\partial A_p(x_s)$ of a stable equilibrium $x_s$ is called a decomposition point. However, not all type-one equilibrium points are decomposition points. Assume that $c = f(x_d)$, as the objective function value increases from $c - \varepsilon$ to $c + \varepsilon$, the number of path components of the level set $S_c := \{x \in M: f(x) < c\}$ decreases by one. For $x_1$ which is a type one equilibrium point but not a decomposition point, when the objective function value increases from $f(x_1) - \varepsilon$ to $f(x_1) + \varepsilon$, the number of path components of $S_c$ remains unchanged (Jongen et. al., 1995; Chiang and Ahmed, 1996).

Suppose that $x$ is a local minimum of $f(x)$ and $\lambda_1(x) \leq \lambda_2(x), \ldots, \lambda_n(x)$ be the eigenvalues of $\nabla^2 f(x)$ and suppose that $v_j(x)$ is normalized eigenvector associated with $\lambda_j(x)$. Since $\nabla^2 f(x)$ is symmetric, $P_j(x) = v_j(x)v_j^T(x)$ is the orthogonal projection matrix on the eigenspace associated with $\lambda_j(x)$. Using the spectral factorization theorem, $\nabla^2 f(x)$ can be rewritten as

$$\nabla^2 f(x) = \sum_{j=1}^{n} \lambda_j(x) P_j(x) \quad (10)$$

Define $\Gamma_i(x), i = 1, \ldots, n$ as

$$\Gamma_i(x) = \sum_{j=1}^{i} P_j(x) - \sum_{j=i+1}^{n} P_j(x) \quad (11)$$

Define the $i$-th order reflected gradient vector field $\Theta_i(x)$

$$\Theta_i(x) = \Gamma_i(x) \nabla f(x) \quad (12)$$

Next, we present the algorithm to find decomposition points on $\partial \bar{A}(x)$ using the reflected gradient method and store them in $\Omega$.

*Algorithm for finding decomposition points:*

*Step 1: Initialization*
    *Step 1.1*: Choose $q$ points $\varrho_j$, $j = 1, \ldots, q$ from a neighborhood of the local minimum $x$ of $f(x)$
    *Step 1.2*: Set a sufficiently small number $\epsilon$
    *Step 1.3*: Set $\Omega = 0$;

*Step 2: Find decomposition points*
**for** $j = 1$ to $q$ do:
    *Step 2.1*: Integrate $\Theta_1(x)$ with the initial condition $\varrho_j$ until $\|\Theta_1(x)\|$ reaches a local minimum, say $\varrho_j^*$, or $\|\Theta_1(x)\|$ becomes unbounded.
    *Step 2.2*: Solve the algebraic equation $f(x) = 0$ using initial condition $\varrho_j^*$ and find solution $x_j$.
    *Step 2.3*: Check the type of equilibrium point with respect to $-\nabla f(x)$:
  **if** $x_j$ is type-one equilibrium point:
    Compute the unstable eigenvector $\vartheta^u$ of $J_{-\nabla f(x)}(x_j)$;
    Numerically integrate PGS with initial points $x_j - \epsilon \vartheta^u$ and $x_j + \epsilon \vartheta^u$ until it approaches the equilibrium points $x_1$ and $x_2$.
    **if** $x_1 \neq x_2$, $\Omega = \Omega \cup \{x_j\}$;
    endif;
  endif;
**end**.

The algorithm finds two local minima, $x_1$ and $x_2$, associated with $x_j$. The efficacy of the algorithm depends to the convergence of the nonlinear solver and the number of initial points chosen around the local minimum $x$. When the objective function has certain periodic properties, upper and lower bounds on the number of initial points can be found which are dependent to the number of objective function variables (Chiang and Chu, 1996).

## 3. Application of trajectory-based optimization to neural network training

In this study, we use a recurrent neural network with one hidden layer. The structure of the network is shown in Fig. 1.

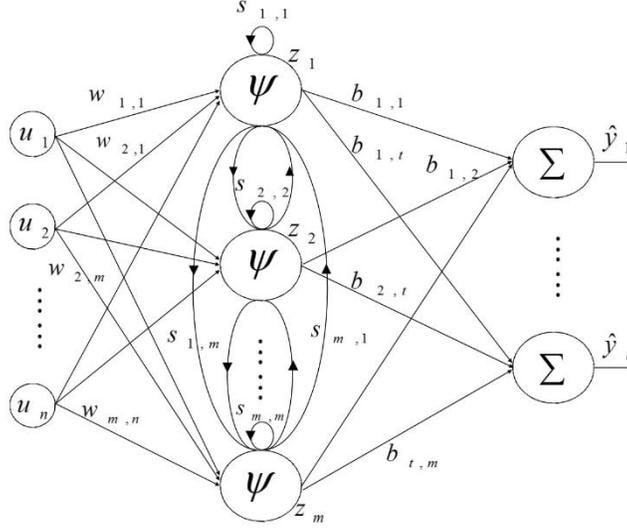

**Fig. 1.** Structure of the recurrent neural network

The input-output relation of the network is

$$\begin{aligned} \mathbf{z}(k) &= \boldsymbol{\psi}\big(W\mathbf{u}(k) + B\mathbf{z}(k-1)\big) \\ \hat{\mathbf{y}}(k) &= V\,\mathbf{z}(k) \end{aligned} \qquad (13)$$

where $\mathbf{u}(k)$ is the input, $\mathbf{z}(k)$ is the internal state, and $\hat{\mathbf{y}}(k)$ is the output of the network. $W$, $B$ and $V$ are neural network weights matrices and $\boldsymbol{\psi}(.)$ is a vector of hyperbolic tangent functions. Assuming that the network has $n$ inputs, $m$ hidden layer nodes and $t$ outputs then $W \in R^{m \times n}$, $B \in R^{m \times m}$, and $V \in R^{t \times m}$. The cost function for finding the optimal network weights is the sum of the squared errors

$$\text{SSE} = \sum_{k=1}^{N} \mathbf{e}(k)^T \mathbf{e}(k) = \sum_{k=1}^{N} \big(\hat{\mathbf{y}}(k) - \mathbf{y}(k)\big)^T \big(\hat{\mathbf{y}}(k) - \mathbf{y}(k)\big) \qquad (14)$$

where $\mathbf{y}(k)$ is the measured output vector, $\hat{\mathbf{y}}(k)$ is the neural network output and $N$ is the number of training samples. The dynamical trajectory-based methodology provides a systematic approach to explore components of feasible region of the optimization problem by repeatedly switching between PGS and QGS. To define the PGS and QGS we need to partition the weight matrices as follows

$$V = \begin{bmatrix} \mathbf{v}_1^T \\ \vdots \\ \mathbf{v}_m^T \end{bmatrix}_{t \times m}, \; W = \begin{bmatrix} \mathbf{w}_1^T \\ \vdots \\ \mathbf{w}_m^T \end{bmatrix}_{m \times n}, \; B = \begin{bmatrix} \mathbf{b}_1^T \\ \vdots \\ \mathbf{b}_m^T \end{bmatrix}_{m \times m} \qquad (15)$$

The constraints are defines as the upper and lower bounds on the network parameters. Define $\mathbf{o}$ as the vector containing all the elements of $W, B$ and $V$

$$\begin{aligned} \mathbf{o} &= [o_i]_{n_p \times 1} = [\mathbf{v}_1, \ldots, \mathbf{v}_m, \mathbf{w}_1, \ldots, \mathbf{w}_m, \mathbf{b}_1, \ldots, \mathbf{b}_m]^T \\ n_p &= m^2 + m \times (n + t) \end{aligned} \qquad (16)$$

The constraints are defined as

$$|o_i| \le l_i, \quad i = 1, \ldots, n_p \qquad (17)$$

The vector of limits on the parameters is defined as $\mathbf{l} = [l_i]_{n_p \times 1}$. By introducing slack variables $\mathbf{s}^T = [s_1, \ldots, s_{n_p}]$, the constraints can be rewritten as

$$h_i(x) = o_i^2 - l_i^2 + s_i^2 = 0, \; i = 1, \ldots, n_p \qquad (18)$$

The vector of optimization problem parameters is defined as

$$\mathbf{x} = [\mathbf{o}\ \mathbf{s}]^T{}_{(2 \times n_p) \times 1} \qquad (19)$$

Since $x$ contains all the elements of $W$, $B$ and $V$, the error, the cost function, and the constraints are functions of $x$

$$\text{SSE} = f(x)$$
$$h(x) = [h_i(x)]_{n_p \times 1} = x_i^2 - l_i^2 + x_{n_p+i}^2 = 0, i = 1, \ldots, n_p \tag{20}$$

Then the problem of finding optimal values of network weights can be written as a constrained optimization problem

$$\min f(x)$$
$$\text{s.t. } h(x) = 0 \tag{21}$$

$Dh$ is defined as

$$Dh(x) = [\partial h_i(x)/\partial x]^T_{(n_p) \times (2 \times n_p)}$$
$$\partial h_i(x)/\partial x = [0, \ldots, 0, 2x_i, 0, \ldots 0, 2x_{i+n_p}, 0, \ldots, 0] \tag{22}$$

Therefore, all the elements of $[\partial h_i(x)/\partial x]$ are zero except for the $i^{th}$ and $(n_p + i)^{th}$ elements. $Dh(x)$ is a full rank matrix and the PGS and QGS for training the neural network are defined as
PGS:

$$\dot{x} = -\left(I - Dh(x)^T (Dh(x)Dh(x)^T)^{-1} Dh(x)\right) \nabla f(x) \tag{23}$$

QGS:

$$\dot{x} = -Dh(x)^T h(x) \tag{24}$$

The process of training neural network using trajectory-based method starts with QGS at an arbitrary initial point. QGS finds a feasible component of feasible region, say $M_i$, and then PGS starts finding local minima in $M_i$. After finding a set of local minima in $M_i$, QGS is invoked again to escape from $M_i$ and move towards another feasible component $M_j$. By repeating this process multiple feasible components of feasible region are located and the algorithm finds a set of local optimal solutions of the optimization problem.

## 4. Stability analysis

The stability of the training method is a critical issue for any training algorithm. In addition, uncertainties and measurement noise may cause the training algorithm to go unstable and must be considered in stability analysis.

**Theorem 1**: The equilibrium points of the QGS are asymptotically stable.

***Proof***: Consider the Lyapunov function $V(x) = h^T(x)h(x)$. $V(x)$ is a locally positive definite function of the state that is equal to zero at global optima of the optimization problem. Thus, $V(x)$ is a locally positive definite function in the vicinity of each equilibrium point. The derivative of the Lyapunov function along the system trajectories is

$$\dot{V} = \left(\frac{\partial V}{\partial x}\right)^T \dot{x} = -h^T Dh Dh^T h = -\|Dh^T h\|^2 \tag{25}$$

The derivative of the Lyapunov function is negative definite in the vicinity of each equilibrium point of the QGS, which are the local minima of the optimization problem. The Jacobian $Dh$ is positive definite in the vicinity of the equilibrium points because they are minima of the cost function. Therefore, all the equilibrium points of the QGS are locally asymptotically stable.

QGS dynamics (24) are independent of the neural network input. Measurements errors can perturb a neural network and drive it outside the basin of attraction of a stable equilibrium during training. However, QGS remains stable in the presence of bounded measurement errors.

If measurement errors cause the neural network input to change from $u$ to $u + \Delta u$, then the PGS will terminate at a perturbed point, which is then used to initialize the QGS. The perturbed QGS initial condition can: (i) lie in the stable region of the previous stable equilibrium point, (ii) lie in the stability region of another stable equilibrium point, or (iii) be infeasible. In case (i), the QGS goes to the previous component of the feasible region and the algorithm continues. In case (ii), the QGS goes to another stable equilibrium point in another feasible component but will return to the current component in future calls of the QGS. In case (iii), the QGS goes to another feasible component. Therefore, measurement noise only affects the order of exploring the components of the feasible region.

Next, we discuss PGS dynamics. This requires the definition of the projection

$$P_r = \left(I - Dh(x)^T (Dh(x)Dh(x)^T)^{-1} Dh(x)\right) \tag{26}$$

$$P_r = [pr_{i,j}], \quad pr_{i,j} = \begin{cases} \alpha_i, & \text{if } i = j,\ 0 \leq \alpha_i \leq 1 \\ \pm\sqrt{1-\alpha_i^2}, & \text{if } |i-j| = n_p \\ 0, \text{elsewhere} \end{cases}$$

Clearly, the norm of the projection matrix is bounded above as $\|P_r\| \leq 1$. The PGS can be rewritten in terms of the projection matrix as

$$\dot{x} = -P_r \frac{\partial}{\partial x}\left(\sum_{k=1}^{N} e(k)^T e(k)\right) = -P_r \sum_{k=1}^{N} \left(\frac{\partial e(k)}{\partial x}\right)^T e(k) \tag{27}$$

We now show that the equilibria of the PGS are stable.

**Theorem 2**: The equilibrium points of the PGS are asymptotically stable.
**Proof**: Consider the Lyapunov function candidate $V(x) = \sum_{k=1}^{N} e(k)^T e(k)$ where $N$ is the number of inputs. Note that since $e(k) = \hat{y}(k) - y(k)$ and $\hat{y}(k)$ is function of $x$, $e(k)$ is a function of $x$. In addition, $e(k)$ is zero only at equilibrium points of the PGS, which are local minima of the optimization problem and is positive elsewhere. Thus, $V(x)$ is a positive definite function of $x$. The derivative of the Lyapunov function along trajectories of PGS is

$$\dot{V} = \dot{x}^T \frac{\partial V}{\partial x} = \dot{x}^T \sum_{k=1}^{N} De(k)^T e(k) = \left[\sum_{k=1}^{N} De(k)^T e(k)\right]^T P_r^T \left[\sum_{k=1}^{N} De(k)^T e(k)\right] \tag{28}$$

where $De(k) = \partial e(k)/\partial x$. Assuming no repeated measurements, $De(k)^T e(k)$ is a full rank matrix and $P_r$ is positive semidefinite, $\dot{V}$ is negative semi definite function of the states and is zero at local optimal solutions of optimization problem and in the null space of $P_r$. $P_r$ is projection matrix on the tangent space of the constraint set and since the equilibrium points are in the constraint set, there is no equilibrium point in the null space of $P_r$. Hence, $\dot{V}$ is negative definite in the feasible region except at the equilibrium points where it is zero. By La Salle's theorem, the equilibrium points of the PGS are locally asymptotically stable.
∎

The input to the neural network is a parameter that determines the PGS dynamics, i.e. $\dot{x} = -c(x, u)$. Therefore, measurement errors can affect the PGS dynamics and make it unstable. If measurement errors change the neural network input from a nominal value $u^*$ to $u^* + \Delta u$, the Taylor series expansion in terms of $\Delta u$ is

$$\dot{x} = -c(x, u) - \frac{\partial c(x, u)}{\partial u}\Delta u + \cdots = -c(x, u) + g(x, u, \Delta u) \tag{29}$$

Suppose that $U$ is subspace of $R^n$ that contains the equilibrium point $x$, i.e. $x \in U \subset R^n$. For a smooth activation function, $g(x, u^*, \Delta u)$ is continuously differentiable and consequently Lipschitz in $R^n$ for all $t \geq 0$. We assume that $g$ satisfies the linear growth bound

$$\|g(x, u^*, \Delta u)\| \leq \gamma \|x\|, \quad \forall t \geq 0, \quad \forall x \in U \tag{30}$$

The next theorem examines the effect of an input perturbation on the PGS.

**Theorem 3**: Given a neural network with $n$ inputs, $m$ hidden layer nodes, $t$ outputs and $N$ measurements. Assume that the network input has the bound $\|u\| \leq k_u$ and the network output has the bound $\|y\| \leq k_y$. If the perturbed PGS dynamics satisfies the linear growth constraint (29), then the equilibrium of the perturbed PGS is asymptotically stable if

$$\gamma < N\sqrt{Nm}\left[\left(\sqrt{m} + \sqrt{k_y(\sqrt{n}k_u + m)}\right)^2\right] \tag{31}$$

**Proof**: Recall that the error $e(k)$ is a function of $x$ and consider the Lyapunov function candidate $V(x) = e(k)^T e(k)$. $V(x)$ is a positive definite function of the states and only becomes zero at optimal solutions of the optimization problem. The derivative of $V(x)$ including the perturbation is

$$\dot{V} = (-e(k)^T De(k) P_r^T + g^T) De(k)^T e(k)$$
$$\leq -\lambda_{\min}(P_r^T) \|De(k)^T e(k)\|^2 + \|g\| \|De(k)^T e(k)\| \tag{32}$$

For negative definite $\dot{V}$, we need the condition

$$\|g\| < \lambda_{\min}(P_r) \|De(k)^T e(k)\| \tag{33}$$

The projection matrix $P_r$ satisfies $\lambda(P_r) \in \{0,1\}$ and the trajectories of the PGS do not pass through its null space (Lee and Chiang, 2004; Lee and Chiang, 2001). Hence, we rewrite (32) as

$$\|g\| < \|De(k)^T e(k)\| \tag{34}$$

Using (13) and the unity upper bound of hyperbolic functions we have the upper bound

$$\|\hat{y}\| \leq Nm\|x\| \tag{35}$$

For any bounded output, we have

$$\|e(k)\| = \|\hat{y} - y\| \leq \|\hat{y}\| + \|y\| \leq N(m\|x\| + k_y) \tag{36}$$

For any bounded input, we have the $\|De(k)^T\|_F$ upper bound

$$\|De(k)^T\|_F \leq \sqrt{N}m(1 + \sqrt{n}\|u\|\|x\| + m\|x\|) \tag{37}$$

We now need the following identity:

$$\forall A \in R^{m \times n}: \|A\|_2 \leq \|A\|_F \leq \sqrt{r}\|A\|_2 \tag{38}$$

where $\|A\|_F$ is the Frobenius norm of $A$ and $r$ is its rank (Meyer, 2000). Using (37) gives

$$\|De(k)^T\|_2 \leq \sqrt{N}m(1 + \sqrt{n}\|u\|\|x\| + m\|x\|) \tag{39}$$

By combining (39), (36) and (34), we get

$$\gamma\|x\| \leq \sqrt{N}m(1 + \sqrt{n}k_u\|x\| + m\|x\|)N(m\|x\| + k_y) \tag{40}$$

For negative definite $\dot{V}$, we require

$$\gamma < N\sqrt{N}m\left[\left(\sqrt{m} + \sqrt{K_y(K_u\sqrt{n} + m)}\right)^2\right] \tag{41}$$

●

## 5. Simulation results

To demonstrate the performance of the dynamical trajectory-based methodology, we apply it to the identification of three nonlinear dynamical systems and compare the results to training using genetic algorithm. While the results for the training data are similar for both approaches, the generalization capability of the dynamical trajectory-based approach is better than that of genetic algorithm and EBP. The results of EBP from (Khodabandehlou and Fadali, 2017) are much worse and are not included for clarity of the figures. We compare to EBP only in the mean squared error tables.

### 5.1. Example 1: NARMA system

The first example is a 10[th] order NARMA system chosen from (Jaeger, 2002). The nonlinear system dynamics is described as

$$y(k+1) = 0.3y(k) + .05y(k)\sum_{i=1}^{9} y(k-i) + 1.5 \times d(k-9)d(k) + 0.1 \tag{42}$$

with input $d(k)$ and output $y(k)$. The input is random zero-mean normally distributed with standard deviation $\sigma = 0.5$. Thus, the input signal is persistently exciting. The target value for neural network training is $y(k+1)$ and the neural network input is $u_n(k) = [d(k), \dots, d(k-9), y(k), \dots, y(k-4)]^T$. $N = 150$ training samples were created, of which 100 samples were used as the training data and 50 samples were used to test the generalization capability of the neural network. All the network parameters were initialized randomly with zero-mean normal values of standard deviation $\sigma = 1$. The optimal number of hidden layer nodes was found by plotting the generalization error versus the number of hidden layer nodes and was found to be $m = 6$. The activation function of the hidden layer nodes is

$$\psi(x) = \tanh(x) = \frac{\exp(x) - \exp(-x)}{\exp(x) + \exp(-x)} \tag{43}$$

The weights in $V$ were constrained to the interval $[-10,10]$, while the elements of $W$ and $B$ were in the interval $[-5,5]$. The limits on elements of $W$ and $B$ are smaller to keep the hidden layer neurons active. To find decomposition points, 30 initial points around the local minimum were created by adding a vector of zero-mean normal random values with standard deviation $\sigma = 0.01$ and $\epsilon$ for finding the decomposition points is chosen to be 0.01.

After invoking the PGS and QGS algorithms repeatedly, we found 6 components of the feasible region and 47 local minima.

Fig. 2 shows the output of the system and the networks for training data and Fig. 3 shows the output of the system and the networks for test data. Fig. 4 shows the generalization error for both neural networks. Both Fig. 3 and Fig.4 show that the trajectory-based trained network can learn the dynamic of the system better than the genetic algorithm trained network. When the output of the system is very close to zero, a small generalization error leads to a large generalization error percent. This can be seen in time steps such as $k = 10$ and $k = 27$ in Fig. 4. Excluding this outlier points, the maximum generalization error for dynamical trajectory-based trained network is 7.4% while the maximum generalization error of genetic algorithm trained network is 22%, which again indicates the superior performance of the trajectory-based approach. Table 1 shows the average mean squared generalization error for different random test data sets. The table shows that dynamical trajectory-based training outperforms genetic algorithm and EBP trained networks by a large margin.

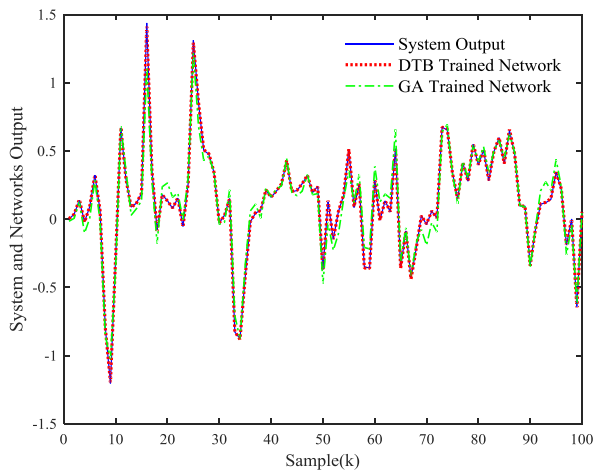

**Fig. 2.** The output of the system and networks for training data. DTB stands for dynamical trajectory-based and GA stands for genetic algorithm

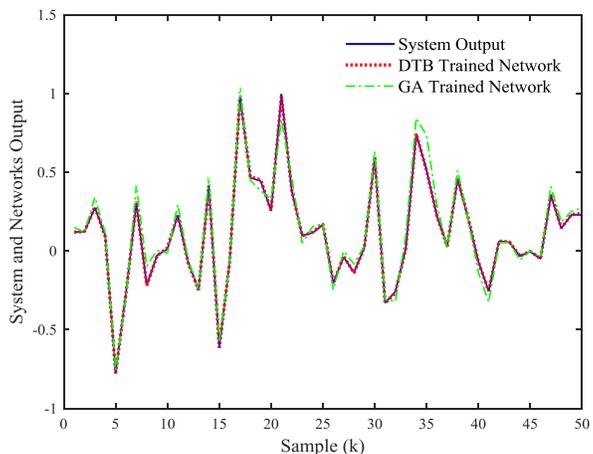

**Fig. 3.** The output of the system and networks for test data

**Table 1**
Mean squared error

| Train Method | DTB | GA | EBP |
|---|---|---|---|
| MSE | 0.0461 | 0.0715 | 0.1184 |

In (Khodabandehlou and Fadali, 2017), another trajectory-based method was used to train this recurrent neural network to identify the NARMA system. The method uses QGS trajectories but the QGS dynamics is different from the one used in this study. With the same input and number of nodes and number training samples, the dynamical trajectory-based method has better generalization performance than that of (Khodabandehlou and Fadali, 2017), and reduces the maximum generalization error from 12.25 to 7.4%.

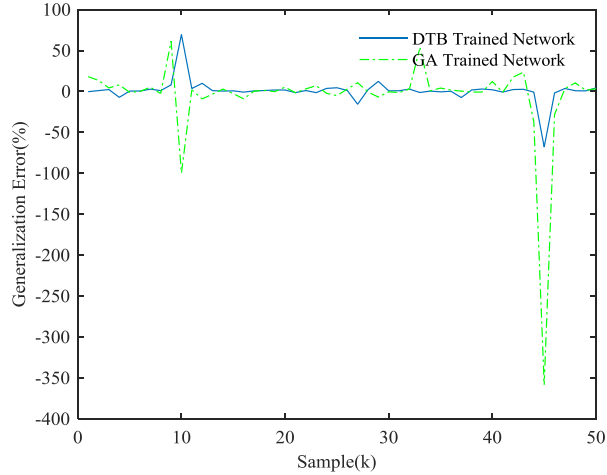

**Fig. 4.** Generalization error for test data

*5.2. Example 2: nonlinear second order system*

Our second example is a second order nonlinear system from (Atiya and Parlos, 2000). The system is governed by the equation

$$y(k+1) = \frac{y(k)y(k-1)(y(k)+0.25)}{1+y(k)^2+y(k-1)^2} + d(k) \tag{44}$$

The network input vector is chosen as $\boldsymbol{u}_n(k) = [d(k), d(k-1), y(k), y(k-1)]^T$ and the number of hidden layer nodes is $m = 8$. As in Example 1, the training set is created using 100 zero-mean normal distributed samples and the test data is created with 50 samples from the same distribution. The PGS and QGS phases are invoked repeatedly to find local minima of the optimization problem. QGS locates 5 feasible components and PGS locates 41 local minima in the feasible components. We used the same values as in Example 1 for the number of initial points for finding decomposition points and $\epsilon$. The local minimum with the best generalization capability is chosen as the optimal solution of the optimization problem.

Fig. 5 shows the output of the system and the networks for training data. It shows that the dynamical trajectory-based trained network is able to learn the behavior of the system better than error genetic algorithm trained network. Fig. 6 shows the output of the system and the networks for test data and illustrates the superior generalization performance of the dynamical trajectory-based trained network. Fig. 7 shows the generalization error for both neural networks. In some samples, the output of the system is close to zero, in such points even small generalization error will lead to big generalization error percentage such as $k = 32$ and $k = 38$ in Fig. 7. Excluding outliers, the maximum generalization error of the dynamical trajectory-based method is 13.3% while the maximum generalization error of the genetic trained network trained network is 31.2%.

Table 2 shows the average mean squared generalization error for different random test data sets. The table shows that dynamical trajectory-based training outperforms genetic algorithm trained network by a large margin. Both the trajectory-based approach and genetic algorithm training have superior performance to EBP trained network.

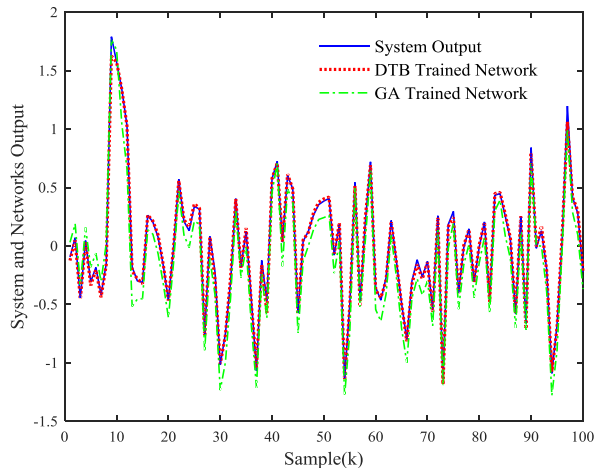

**Fig. 5.** The output of the system and networks for training data.

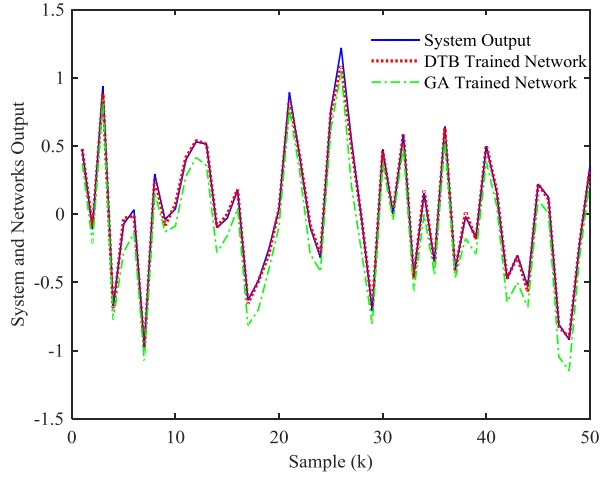

**Fig. 6.** The output of the system and networks for test data

**Table 2**
Mean squared error

| Train Method | DTB | GA | EBP |
|---|---|---|---|
| MSE | 0.0523 | 0.0724 | 0.1297 |

In (Khodabandehlou and Fadali, 2017), we used another trajectory-based method to train a recurrent neural network to identify the same system. With the same number of hidden nodes and the same input and fewer training samples, the dynamical trajectory-based method has better generalization performance and reduces the maximum generalization error from 18.7% to 13.3%.

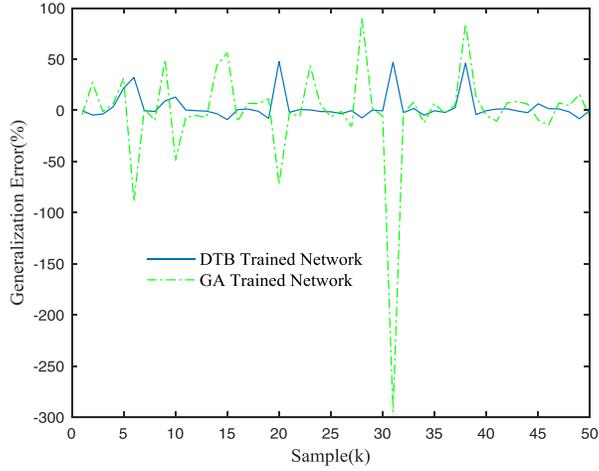

**Fig. 7.** Generalization error for test data

*5.3. Example 3: Bouc-Wen System*

Bouc-Wen model is a nonlinear hysteretic system that has been widely studied in mechanical and civil engineering (Ismail et. al. (2009); Khodabandehlou et. al. 2017). Identification of Bouc-Wen system is a challenging problem due to nonlinearity and hysteretic behavior of the model. The governing equation of the model is described as

$$m_L \ddot{y}(t) + r(y, \dot{y}) + z(y, \dot{y}) = s(t) \tag{45}$$

$s(t)$ is the input, $y(t)$ is the output and $m_L$ indicates the mass constant of the system. $r(y, \dot{y})$ represents the restoring force and $z(y, \dot{y})$ is nonlinear term which cipher the hysteretic behavior of the model. The restoring force is governed by

$$r(y, \dot{y}) = k_L y + c_L \dot{y} \tag{46}$$

$k_L$ and $c_L$ are linear stiffness and viscous damping coefficients respectively. $z(y, \dot{y})$ is described by the first order differential equation

$$\dot{z}(y, \dot{y}) = \alpha |\dot{y}| - \beta(\gamma |\dot{y}||z|^{v-1} + \delta \dot{y}|z|^v) \tag{47}$$

$\alpha, \beta, \gamma, \nu$ and $\delta$ are Bouc-Wen parameters. These parameters determine the shape and smoothness of the hysteresis loop. Table 3 shows the values of these parameters (Noël and Schoukens, 2016).

Table 3
Bouc-Wen model parameters

| Parameter | $m_L$ | $c_L$ | $k_L$ | $\alpha$ | $\beta$ | $\gamma$ | $\delta$ | $\nu$ |
|---|---|---|---|---|---|---|---|---|
| Value | 2 | 10 | $5*10^4$ | $5*10^4$ | $10^3$ | 0.8 | -1.1 | 1 |

For simulation, the network parameters were initialized with random variables from a zero-mean normal distribution with standard deviation of $\sigma^2 = 0.1$. The optimal number of hidden layer nodes was found to be $m = 7$. The input vector to the neural network is $u(k) = [s(k), \ldots, s(k-5), y(k-1), \ldots, y(k-5)]$ and the target output for training neural network is $y(k)$. We use the same constraints on neural network weights as Example 1 to find the decomposition points and $\epsilon$. We choose the local minimum with the best generalization performance as the global minimum of the optimization problem. The dynamical trajectory-based method, finds 6 components of the feasible region and 27 local minima.

Fig. 8 shows the output of the Bouc-Wen system and output of the trained networks for training data and Fig. 9 shows the output of the networks for test data. Although the performance of the networks is close, the trajectory-based network has slightly better performance on both training and test data. Fig. 10 clearly depicts the better performance of the trajectory-based network. Excluding the points that output of the system is very close to zero, the maximum generalization error of the genetic algorithm trained network is two times larger than that of the trajectory-based trained network. Table 4 summarizes the mean squared error of the identification of Bouc-Wen system for test data. The results show the superior performance of the trajectory-based network.

Table 4
Mean squared error

| Train Method | DTB | GA | EBP |
|---|---|---|---|
| MSE | 0.00161 | 0.00274 | 0.0139 |

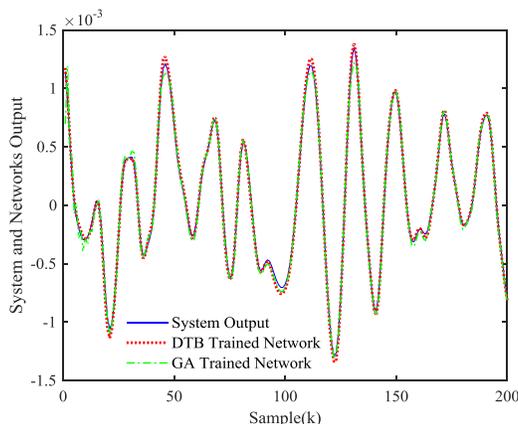

**Fig. 8.** The output of the system and networks for training data.

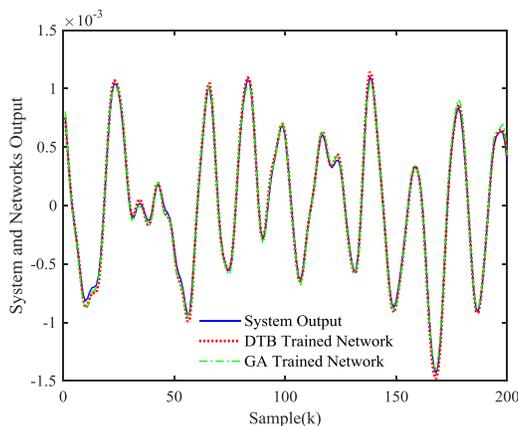

**Fig. 9.** The output of the system and networks for test data

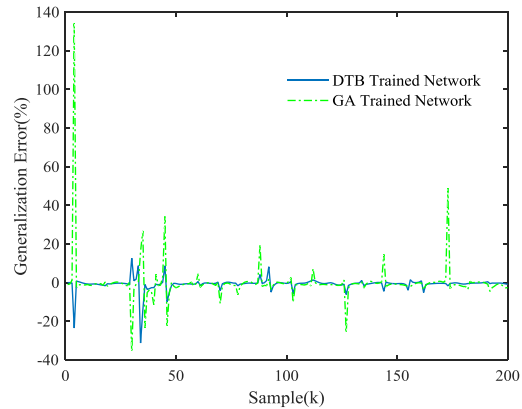
**Fig. 10.** Generalization error for test data

The performance of the algorithm depends on three factors: a) The upper and lower bound constraints on the network weights, b) the number of initial points used to find decomposition points, and c) $\epsilon$ for finding initial points.

The upper and lower bounds can be chosen arbitrary large to avoid the effect of the constraints on the final solution. With large bounds, the algorithm searches larger space for components of the feasible region. Because of the simple structure of the QGS that the algorithm uses to find connected components, searching larger space does not increase the training time significantly. The number of initial points for finding decomposition points is another factor that affects the performance of the algorithm. A larger number of initial points improves the performance of the algorithm but increases the training time. In our simulations, we found that using $0.1 \times n_p$ to $0.2 \times n_p$ initial points is enough to get satisfactory results. Choosing a very small value for $\epsilon$ can prevent the algorithm from finding decomposition points, which play a pivotal role in finding locally optimal solutions. Based on our simulation, a good range for $\epsilon$ is $\in [0.01, 0.1]$.

## 6. Conclusion

This study uses a dynamical trajectory-based methodology for training artificial neural networks. The methodology provides a systematic approach for finding multiple solutions of general nonlinear optimization problems. By invoking PGS and QGS phases repeatedly, the algorithm is able to locate multiple feasible components of feasible region and locate the local minima in the feasible components. Lyapunov theory was used to prove the stability of the method and its stability in presence of measurement error. Calculating PGS has a similar computational cost to calculating the derivative of the cost function with respect to network variables and is therefore similar to calculations of the EBP algorithm. Finding QGS is relatively easy because the constraints on the network weights are simple. Thus, the effort required to implement the algorithm is comparable to the effort required to implement EBP. There has to be a tradeoff between the number of initial point for finding decomposition points and the training time of the algorithm. Increasing the number of initial points enhances the performance of the algorithm but increases the total training time. Proper choose of $\epsilon$ is another factor that affects the performance of the method. Choosing very small values for $\epsilon$ may prevent the method from finding local minima. The training time of the algorithm is comparable to global optimization approaches and is slower than Newton-based methods. Therefore, applying this method to deep or convolutional neural networks will require GPU processing  Simulation results show that the dynamical trajectory-based method provides better performance than genetic algorithm and EBP trained networks. Future work will modify the algorithm to learn both the structure and the internal weights of the neural network.

## References

Abass, H. A. (2003). Speeding up back-propagation using multi-objective evolutionary algorithms. *Neural Computation*, 15, 2705–2726.
Atiya, A. F., & Parlos, A. G. (2000). New results on recurrent network training: Unifying the algorithms and accelerating convergence. *IEEE Transactions on Neural Netwo*rks, 11(3), 697–709.
Blanco, A., Delgado, M., & Pegalajar, M. C. (2001). A real-coded genetic algorithm for training recurrent neural networks. *Neural Networks*, 14, 93-105.
Cai, X., Venayagamoorthy, G. K., & Wunsch, D. C. (2010). Evolutionary swarm neural network game engine for Capture Go. *Neural Networks*, 23(2), 295-305.
Chiang, H. D., & Ahmed, L. F. (1996). Quasistability regions of nonlinear dynamical systems: optimal estimation. *IEEE Transactions on Circuits and Systems*, 43, 636-643.
Chiang, H. D., & Chu, C. C. (1996). A systematic search method for obtaining multiple local optimal solutions of nonlinear programming problems. *IEEE Transactions on Circuits and Systems I: Fundamental Theory and Applications, 43(2), 99-109.*
Fu, X., Li, S., Fairbank, M., Wunsch, D. C., & Alonso, E. (2015). Training recurrent neural networks with the Levenberg-Marquardt algorithm for optimal control of a grid-connected converter. *IEEE Transactions on Neural Networks and Learning Systems*, 26(9), 1900-1912.
Georgiopoulos, M., Li C., & Kocak, T. (2011). Learning in the feed-forward random neural network: A critical review. *Performance Evaluation*, 68(4), 361–384.
Ghazikhani, A., Akbarzadeh, T. M. R., & Monsefi, R. (2011). Genetic regulatory network inference using recurrent neural networks trained by a multi agent system. In *Proceedings of 1st International eConference on Computer and Knowledge Engineering,* (ICCKE), (pp. 95-99), Mashhad, Iran.
Gori, M., & Tesi, A. (1992). On the problem of local minima in backpropagation. *IEEE Transactions on Pattern Analysis and Machine Intelligence*, 14(1), 76–86.
Guddat, J., Vazquez, F. G., & Jongen, H. T. (1990). *Parametric optimization: singularities, path following and jumps*. New York, Wiley.

Hermans, M., Dambre, J., & Bienstman, P. (2015). Optoelectronic systems trained with backpropagation through time. *IEEE Transactions on Neural Networks and Learning Systems*, 26(7), 1545-1550.

Hornik, K. (1991). Approximation capabilities of multilayer feedforward networks. *Neural Networks*, 4(2), 251–257.

Jaeger, H. (2002). Adaptive nonlinear systems identification with echo state networks. In *Advances in Neural Information Processing Systems 15 (NIPS 2002)*, (pp. 593-600), MIT Press, Cambridge, MA, USA.

Ismail, M., Ikhouane, F., & Rodellar, J. (2009). The hysteresis Bouc-Wen model, a survey. *Archives of Computational Methods in Engineering*, 16, 161-168.

Jaganathan, S., & Lewis, F. L. (1996). Identification of nonlinear dynamical systems using multilayered neural networks. *Automatica*, 32(12), 1707-1712.

Jongen, H. T., Jonker, P., & Twilt F. (1995). *Nonlinear Optimization in $R^n$*. Germany, Frankfurt: Peter Lang Verlag.

Khodabandehlou, H., & Fadali, M. S. (2017). A Quotient Gradient Method to Train Artificial Neural Networks. In *Proc. Int. Joint Conf. Neural Networks* (IJCNN 2017).

Khodabandolehlou, H., Pekcan, G., Fadali, M. S., Salem, M.A. (2017). Active neural predictive control of seismically isolated structures. *Structural Control and Health Monitoring*, 25(1), doi: https://doi.org/10.1002/stc.2061

Kiranyaz, S., Turker, I., Yildirim, A., & Gabbouj, M. (2009). Evolutionary artificial neural networks by multi-dimensional particle swarm optimization. *Neural Networks*, 22, 1448–1462.

Luitel, B., & Venayagamoorthy, G. K. (2009). A PSO with quantum infusion algorithm for training simultaneous recurrent neural networks. In *Proceedings of International Joint Conference on Neural Networks,* (IJCNN), (pp. 1923-1930), GA, USA.

Lee, J., & Chiang, H. D. (2001). Computation of multiple type-one equilibrium points on the stability boundary using generalized fixed-point homotopy methods. In *Proceedings of IEEE International Symposium on Circuits and Systems*, (pp. 361-364), Sydney, NSW, Australia.

Lee, J., & Chiang, H. D. (2004). A dynamical trajectory-based methodology for systematically computing multiple optimal solutions of general nonlinear programming problems. *IEEE Transactions on Automatic Control*, 49(6), 888-899.

Martens, J., & Sutskever, I. (2011). Learning recurrent neural networks with hessian-free optimization. *In Proceedings of the 28th International Conference on Machine Learning* (ICML-11), 1033–1040.

Meyer, C. (2000). *Matrix Analysis and Applied Linear Algebra*. Society for Industrial and Applied Mathematics, Philadelphia, PA, USA.

Mirikitani, D. T. & Nikolaev, N. (2007). Recursive Bayesian Levenberg-Marquardt training of recurrent neural networks. In *Proceedings of International Joint Conference on Neural Networks*, (IJCNN), (pp. 1089–1098), Orlando, FL, USA.

Mirikitani, D. T., & Nikolaev, N. (2010). Recursive Bayesian recurrent neural networks for time-series modeling. *IEEE Transactions on Neural Networks*, 21(2), 262-274.

Monner, D., & Reggia J. A. (2012). A generalized LSTM-like training algorithm for second-order recurrent neural networks. *Neural Networks*, 25, 70-83.

Montingy, S. D. (2014). New approximation method for smooth error backpropagation in a quantron network. *Neural Networks*, 60, 84-95.

Noël, J. P., Schoukens, M. (2016). Hysteretic benchmark with a dynamic nonlinearity. In *Workshop on Nonlinear System Identification Benchmarks*, (pp. 7–14), Brussels, Belgium, Available http://nonlinearbenchmark.org/#BoucWen

Rubio, J. J., & Yu, W. (2007). Neural networks training with optimal bounded ellipsoid algorithm. In *Proceedings of the 4th International Symposium on Neural Networks: Advances in Neural Networks*, (pp. 1173-1182), Nanjing, China.

Salerno, J. (1997). Using the particle swarm optimization technique to train a recurrent neural model. In *Proceedings of the 9th International Conference on Tools with Artificial Intelligence*, (pp. 45–49), Newport Beach, CA, USA.

Sharma, A. A. (2016). Univariate short term forecasting of solar irradiance using modified online backpropagation through time. *2016 International Computer Science and Engineering Conference,* (ICSEC), (pp. 1-6), Chiang Mai, Thailand.

Stroeve, S. (1998). An analysis of learning control by backpropagation through time. *Neural Networks*, 11(4), 709-721.

Tang, Z., Wang, D., & Zhang, Z. (2016). Recurrent neural network training with dark knowledge transfer. *IEEE International Conference on Acoustics, Speech and Signal Processing,* (ICASSP), (pp. 5900-5904), Shanghai, China.

Werbos, P. J. (1998). Generalization of backpropagation with application to a recurrent gas market model. *Neural Networks*, 1, 339-356.

Yu, W., & Rubio J. J. (2009). Recurrent neural networks training with stable bounding ellipsoid algorithm. *IEEE Transactions on Neural Networks*, 20(6), 983–991.

Yu, Y., Xi, L., &, Wang, S. (2007). An improved particle swarm optimization for evolving feed-forward artificial neural networks. *Neural Processing Letters*, 26, 217–231.

Zhang, L., & Suganthan, P. N. (2016). A survey of randomized algorithms for training neural networks. *Information Sciences*, 364–365, 146–155

Zhu, Y. Q., Xie, W. F., & Yao, J. (2006). Nonlinear system identification using genetic algorithm based recurrent neural networks. In *Proceedings of Canadian Conference on Electrical and Computer Engineering,* (CCECE'06)*,* (pp. 571-575), Ottawa, ON, Canada.